# FeAs-based superconductivity: a case study of the effects of transition metal doping on $BaFe_2As_2$.

**Paul C. Canfield and Sergey L. Bud'ko**

**Ames Laboratory, US DOE and Department of Physics and Astronomy**
**Iowa State University**
**Ames, Iowa 50011, USA**

## Abstract:

The recently discovered FeAs-based superconductors are a new, promising set of materials for both technological as well as basic research. They offer transition temperatures as high as 55 K as well as essentially isotropic and extremely large upper, superconducting critical fields in excess of 40 T at 20 K. In addition they may well provide insight into exotic superconductivity that extends beyond just FeAs-based superconductivity, perhaps even shedding light on the still perplexing CuO-based high-$T_c$ materials. Whereas superconductivity can be induced in the RFeAsO (R = rare earth) and $AEFe_2As_2$ (AE = Ba, Sr, Ca)) families by a number of means, transition metal doping of $BaFe_2As_2$, e.g. $Ba(Fe_{1-x}TM_x)_2As_2$, offers the easiest experimental access to a wide set of materials. In this review we present an overview and summary of the effect of TM doping (TM = Co, Ni, Cu, Pd, and Rh) on $BaFe_2As_2$. The resulting phase diagrams reveal the nature of the interaction between the structural, magnetic and superconducting phase transitions in these compounds and delineate a region of phase space that allows for the stabilization of superconductivity.

## Related Resources:

preprint archive (Condensed Matter): http://arxiv.org/archive/cond-mat

Novel Materials and Ground States (at Ames Laboratory, US DOE and Department of Physics and Astronomy, Iowa State University) research group web-page: http://www.cmpgroup.ameslab.gov/personnel/canfield/index.html

*Physica C*, Volume 469, Issues 9-12, Pages 313-674 (1 May 2009-15 June 2009) Special Issue: Superconductivity in Iron-Pnictides

*New Journal of Physics*, Volume 11, February 2009 Special Issue: Focus on Iron Based Superconductors

Journal of the Physical Society of Japan, Online Collection of papers on Iron-Pnictide and Related Superconductors: http://www.ipap.jp/articles/showArticle.cgi?sec=Fe

## Key Terms:

$T_c$ (Critical temperature): The temperature below which a material becomes superconducting in zero applied magnetic field.

$H_{c2}$ (Upper superconducting critical field): The highest applied magnetic field at which a compound still superconducts; is a function of temperature and can be anisotropic, i.e. depends on direction of applied field with respect to crystal lattice.

$j_c$ (The superconducting critical current density): The maximum current density a superconductor can carry without reverting to the non-superconducting state; is a function of temperature as well as magnetic field.

Structural phase transition: A transition from one crystallographic structure to another, e.g. from a tetragonal to an orthorhombic unit cell. Takes place at a characteristic temperature $T_s$.

Antiferromagnetic phase transition: A transition from a state with no long range magnetic order to one with a well defined magnetic periodicity but with no net ferromagnetic component. Takes place at a characteristic temperature $T_N$.

## Introduction

Humanity's quest for materials with better properties is so integral to the fabric of its history that epochs are named after the materials that define them: stone, bronze, iron, and silicon. Currently humanity is desperately trying to develop materials that will allow for improved generation, transport, and even storage of energy. Among these materials, superconductors are held up as having great promise, but also great difficulties that need to be overcome. The three key properties that need to be improved are the superconducting transition temperature, $T_c$, the upper superconducting critical magnetic field, $H_{c2}$, and the critical current density, $j_c$. These properties define the maximum operating temperature, field, and current of the superconductor. Whereas members of the high-$T_c$ copper-oxide materials class become superconducting as high as 130 K, they have highly anisotropic $H_{c2}$ values (with the lower value being quite low), and great (and expensive) effort has be made to get reasonable $j_c$ values (1, 2). Over the past decade, a general feeling emerged that easier to make materials, with more isotropic $H_{c2}$ values may be more useful, even if they had lower $T_c$ values (3).

For its first half century, superconductivity was thought to be the purview of a few rare compounds made up of metallic elements with electron-phonon coupling governing (and BCS theory describing) the salient physics. Over the past thirty or so years, it has been becoming increasingly clear that superconductivity is a very robust ground state and appears to be able to emerge out of a variety of coupling mechanisms with a variety of gap symmetries. The theoretical frameworks that have been developed in efforts to explain superconductivity in all of its manifestations have influenced and advanced not only far flung fields in condensed matter physics, but also astrophysics and current models of the quark ion plasma. With each discovery of new classes of superconductivity theory is provided with experimental benchmarks that secure the expanding intellectual borders of our understanding of this now apparently ubiquitous, low temperature ground state.

During the past 15 years there have been a series of discoveries of superconducting compounds that essentially bridge the gap between the original, low-$T_c$ elements and alloys and the high-$T_c$ copper oxides. In 1994 the isostructural $RNi_2B_2C$ and $YPd_2B_2C$ compounds were discovered with $T_c$ values as high as 17 K and 23 K respectively [cava refs from physics today]. These compounds demonstrated great versatility and also rivaled the existing maximum $T_c$ values for intermetalic compounds (4, 5). Based on the richness of the $RNi_2B_2C$ families many research groups returned to searching for superconductivity in compounds rich in light, often covalently bonding elements. In early 2001 the binary compound, $MgB_2$ was found to superconduct with a $T_c \sim 40$ K (6-8). When doped with carbon $MgB_2$ can have a large and increasingly less anisotropic $H_{c2}$ (7, 8). Carbon doped $MgB_2$ has intermediate $j_c$ values, but these can be achieved by simple and cheap methods; it is currently being extensively studied as a potential, practical superconductor.

Based on $RNi_2B_2C$, $YPd_2B_2C$, and $MgB_2$, searches for new superconductors focused more and more on compounds bearing light elements as well as compounds with more directional, or less metallic, bonding. In February of 2008 the discovery of a new type of superconductor was published; $LaFeAs(O_{1-x}F_x)$ was found to superconduct at temperatures as high as 26 K (9). This discovery was followed by work that showed that the same compound could be made to superconduct at temperatures as high as 43 K under applied pressure (10) and that even higher superconducting transition temperatures (as high as ~55 K) could be found in other (more compact) rare earth members of the $RFeAs(O_{1-x}F_x)$ family (11). These results lead to an extensive, world wide, effort to understand and categorize this apparently new type of superconductor.

A second key discovery was posted in May of 2008; $(Ba_{1-x}K_x)Fe_2As_2$ was found to superconduct at temperatures approaching 40 K (12). On one hand, the parent compounds LaFeAsO (1111) and $BaFe_2As_2$ (122) both shared similar structural features: a square planar Fe layer that is capped above and below by As layers (see Figure 1a). On the other hand, the $BaFe_2As_2$ compounds did not contain any oxygen. These two observations led to an early understanding that the FeAs layer was the key structural motif in these materials and that the superconductivity in the RFeAsO materials was not uniquely associated with oxide physics (as is the case for the high $T_c$ copper-oxide materials).

The discovery of $(Ba_{1-x}K_x)Fe_2As_2$ was important for another, practical, reason. Being composed solely of metallic and metalloid elements $(Ba_{1-x}K_x)Fe_2As_2$ could be readily grown using conventional intermetallic, solution growth techniques. Single crystals of $BaFe_2As_2$, $SrFe_2As_2$ as well as the previously unknown $CaFe_2As_2$ could be grown out of Sn flux (13-18). Single crystals of $(Ba_{0.55}K_{0.45})Fe_2As_2$ with dimensions as large as 4x4x0.2 $mm^3$ (Figure 1b) were grown and characterized. As shown in Figure 2 (19) the $H_{c2}(T)$ plot, determined up to field in excess of 60 T, indicates that these compounds show great initial promise of being practical superconductors. At 20 K, $(Ba_{0.55}K_{0.45})Fe_2As_2$ has a nearly isotropic $H_{c2}$ with the lowest value being in excess of 40 T (a $H_{c2}$ value that is already comparable to the highest $H_{c2}$ possible for fully optimized $Nb_3Sn$) (19, 20).

Although single crystals of K-doped $BaFe_2As_2$ could be grown, controlling the homogeneity of K throughout the crystal was difficult. Wavelength dispersive X-ray spectroscopy (WDS) studies revealed that there could be substantial variation of doping level throughout a single crystal (13) and that different layers of crystal could have different $T_c$ values (21).

A third, important discovery was posted in July of 2008; superconductivity could be stabilized in both LaFeAsO as well as in $BaFe_2As_2$ by substitution of Co for Fe (22, 23). These results were very important for both operational as well as conceptual reasons. Operationally, cobalt and, as will be shown, other transition metals, can be substituted for Fe with far greater ease and reliability than substituting the volatile and reactive K for Ba. Conceptually, the fact that direct substitution into the Fe-layer with another transition metal could stabilize superconductivity made FeAs based superconductivity very different from the copper-oxide-based high-$T_c$ compounds which are notoriously sensitive to perturbations of the Cu sublattice. Over the past year a substantial effort has been invested in delineating and understanding the effects of transition metal substitution into $BaFe_2As_2$, as well as the other alkali-earth compounds, $SrFe_2As_2$ and $CaFe_2As_2$.

Given that research on Fe-based superconductors is in its infancy, a standard, critical review of the field is premature. On the other hand, it is important to provide an overview of what currently is the best studied and understood model system: transition metal doped $BaFe_2As_2$. Such an overview will provide both a sense of the excitement and promise that these materials offer as well as a starting point to researchers interested in learning more about these compounds. To this end, we will outline the current understanding of the interaction between superconductivity, magnetism, crystal structure, and electronic structure in pure and doped $BaFe_2As_2$. Given the nature of this review, as well as its limited space, we will place an emphasis on research we are personally involved with, but, with this acknowledged, we will also try to refer to the appropriate key results in the field. To accomplish this we will place an emphasis on the results of basic structural, thermodynamic and transport measurements. Given the intimate coupling between structure (electronic, crystallographic, and magnetic) and superconductivity, results of X-ray and neutron scattering as well as angle resolved photo emission will also be reviewed. Although there is also a growing literature of NMR and optical studies that can be referred to by interested readers (24-50) these will not be covered in this article.

In reading this review it should be kept in mind that superconductivity can be induced in $BaFe_2As_2$ (as well as other 122 compounds) by substitution on the alkali earth site (as mentioned above), on As site, as well as by application of pressure (hydrostatic as well as non-hydrostatic), and the basic form of the $T$ - $x$ and $T$ - $P$ phase diagrams are remarkably similar. This implies that the basic physics revealed by transition metal doping of this system may well be more generic than simply "adding extra electrons by doping". Indeed the reason for studying $Ba(Fe_{1-x}TM_x)_2As_2$ compounds is the ease and variety of perturbation that the various TM dopants offer combined with the hope that the physics revealed will be of a general nature. In other words, there appear to be common correlations between $T_s$, $T_N$, $T_c$ and control parameters such as doping and pressure. By thoroughly studying the effects of TM doping (the control parameter that offers the greatest range), hopefully a universal picture of the salient physics will emerge.

## Behavior of parent compounds

The parent compounds of 1111 and 122 families manifest structural (high temperature tetragonal to low temperature orthorhombic) and antiferromagnetic phase transitions that are either simultaneous and first order (as is the case for the $CaFe_2As_2$, $SrFe_2As_2$ and $BaFe_2As_2$ (15-18, 51)) or in very close proximity (as is the case for some of the lighter rare earth 1111 compounds (see Ref. 52 and references therein)). A particularly clear example of this can be seen in the thermodynamic and transport data taken on $BaFe_2As_2$ and $CaFe_2As_2$ (Figure 3). These transitions correspond to a simultaneous, first order, structural / antiferromagnetic phase transition giving rise to a low temperature, orthorhombic, antiferromagnetic phase (Figure 4a). Figure 4b shows the same sharp jump and approximately 1K thermal hysteresis in the orthorhombic splitting and the integrated intensity of an antiferromagnetic ordering peak (53, 54).

The temperature dependent resistivity of $BaFe_2As_2$ and $CaFe_2As_2$ reveal two competing features associated with these transitions: a loss of scattering and a decrease in the density of states. For $BaFe_2As_2$ these combine to give a small peak (sometimes so small that is it is hard to detect) followed by a sharp drop in resistivity. For $CaFe_2As_2$ the peak is clearer, but it is still followed by a loss of scattering at lower temperatures. TM doping of $BaFe_2As_2$ leads to resistivity data the manifest these two features to varying degrees.

As will be shown below, doping on the Fe site with *3d* and *4d* transitions metals leads to a suppression as well as a separation of these two transitions, with the antiferromagnetic transition being suppressed more rapidly. This observation makes the fact that these transitions appear to be slightly separated in some of the 1111 materials less puzzling and suggests that the 122 and 1111 parent compounds are part of a continuum of behavior, with the 122 materials possibly being slightly more ordered in some manner.

## $Ba(Fe_{1-x}Co_x)_2As_2$

Single crystals of $Ba(Fe_{1-x}Co_x)_2As_2$ are readily grown from a FeAs self flux. (23, 55-57) Elemental Ba is added to a mixture of FeAs and CoAs and crystals as large as 1x1x0.2 $cm^3$ can be grown (see inset to Figure 5). Although a nominal Co concentration can be assigned to the samples based on the Fe:Co ratio in the melt, it is important to establish the actual Co content in the sample. For all of the *3d*- and *4d*-transition metal dopings reviewed here the actual doping concentration has been determined by WDS measurements on several samples, on several locations on each of several cleaves. These data have allowed for an assessment of the doping homogeneity as defined by twice the standard deviation. For Co doping, as well as for the other transition metal doping, the homogeneity is vastly superior to that we were able to achieve for the K-doped $BaFe_2As_2$.

Figure 5 present the normalized resistivity and magnetic susceptibility for representative samples of $Ba(Fe_{1-x}Co_x)_2As_2$ (55). As Co replaces Fe the resistive signature of the combined structural and magnetic phase transition is suppressed and changes shape. In a similar manner, the sharp decrease in magnetic susceptibility that occurs in $BaFe_2As_2$ is broadened and suppressed as Co replaces Fe. At intermediate dopings, while there is still a clear signature of structural and magnetic ordering, superconductivity becomes apparent in both the resistivity and susceptibility data sets. For still higher Co doping levels the resistive and magnetic signatures of structural and magnetic phase transitions are suppressed and the superconducting transition temperature passes through a maximum, dropping with further Co substitution.

These data can be used to construct a $T-x$ phase diagram, but in order to do this it is important to relate features in the thermodynamic and transport measurements to microscopic data. Figure 6 compares the results of neutron scattering measurements with resistivity and magnetic susceptibility data for the $x = 0.047$ sample of $Ba(Fe_{1-x}Co_x)_2As_2$ (58). There are two clear features in each of the $d\rho/dT$ and $d\chi/dT$ data sets. The lower temperature one corresponds to the onset of magnetic scattering intensity associated with the antiferromagnetic order and the higher temperature one occurs at the temperature where a clear loss of extinction feature marks the occurrence of the tetragonal to orthorhombic phase transition (also seen in x-ray scattering data as the onset of a clear splitting of tetragonal peaks**)**. These features also correspond to features in the temperature dependent specific heat (not shown) (55, 57).

Using these criteria for determination of the structural and magnetic phase transitions, thermodynamic and transport data can be used to assemble the $T-x$ phase diagram for $Ba(Fe_{1-x}Co_x)_2As_2$, shown in Figure 7. As Co is added the simultaneous structural and antiferromagnetic phase transition is split as it is suppressed. A clear dome of superconductivity exists at low temperatures both in the antiferromagnetic / orthorhombic phase as well as in the tetragonal phase. Anisotropic $H_{c2}(T)$ data taken for samples on both sides of the superconducting dome, referred to as the underdoped side, for samples with dopings below the value of $x$ needed to produce maximum $T_c$ (also the value of $x$ that suppressed all features of $T_s$ and $T_N$), and the overdoped side of samples with $x$ larger than this value, show that the $H_{c2}$ anisotropy changes as the structure of the superconducting state changes from orthorhombic and antiferromagnetic to tetragonal with no long range magnetic order (55). $H_{c2}(T)$ data taken for near optimal doping (55) (Figure 2) is qualitatively similar to that found for the $(Ba_{0.55}K_{0.45})Fe_2As_2$ shown in the same figure, but with a slightly lower field scale, most likely reflecting the somewhat lower $T_c$ value.

The changing $H_{c2}$ anisotropy (55), magneto-optical measurements (60), neutron and X-ray scattering data (58, 61), as well as recent NMR (62, 63) and μSR (64) results all indicate that the sample is homogeneously superconducting across the whole dome region: i.e. there is no phase separation associated with the superconducting state in the underdoped region. This being said, although superconductivity clearly co-exists with long range antiferromagnetic order on the underdoped side of the superconducting dome, Figure 7 clearly shows that the integrated intensity of magnetic scattering peaks, associated with the antiferromagnetic order, decrease when the sample is cooled through $T_c$. These data give some of the clearest indication that superconductivity interacts, and very likely competes, with antiferromagnetism.

## *3d* doping

In order to determine how general the $T-x$ phase diagram found for Co-doping is, other, higher $Z$, *3d* transition metals were substituted for Fe (65). Single crystalline samples of $Ba(Fe_{1-x}TM_x)_2As_2$, for TM = Ni, Cu, and Co/Cu co-doping, were grown from excess FeAs and the precise concentration of each of the transition metals was again determined via WDS analysis. Figure 8 shows the normalized resistivity for each of these series.

The $Ba(Fe_{1-x}Ni_x)_2As_2$ series manifests resistivity data that is qualitatively similar to that seen for Co-substitution. The structural / magnetic phase transitions are suppressed and split with increasing Ni-substitution and, at intermediate doping levels, superconductivity is stabilized. On the other hand, the $Ba(Fe_{1-x}Cu_x)_2As_2$ series resistivity data do not show any sign of superconductivity, even though the signatures of the structural / magnetic transitions are suppressed and split in a manner very similar to that seen for Co and Ni substitutions.

The lack of superconductivity in the $Ba(Fe_{1-x}Cu_x)_2As_2$ series raises the question of whether Cu is somehow uniquely pernicious to superconductivity or if there is a more subtle or profound difference? A series of $Ba(Fe_{1-x-y}Co_yCu_x)_2As_2$ samples (with $x \sim 0.024$) was studied to address this question. As can be seen in Figure 5, a Co substitution of $x \sim 0.024$ suppresses and separates the structural and magnetic phase transitions, but does not induce superconductivity. Using this as a starting point, Cu substitution for some of the remaining Fe further suppresses the structural / magnetic phase transitions and induces superconductivity. The fact that Cu can be used to induce superconductivity is inconsistent with it being solely pernicious to the superconducting state and indicates that there may well be something a little more subtle influencing the superconductivity in this system.

$T - x$ phase diagrams for Ni, Cu, and Co/Cu can be inferred from thermodynamic and transport data (65) using the same criteria developed for the $Ba(Fe_{1-x}Co_x)_2As_2$ data (55, 58). In Figure 9 the $T - x$ for these three series are plotted along with the $Ba(Fe_{1-x}Co_x)_2As_2$ data from Figure 7. These data show that whereas the structural / magnetic phase transitions in $Ba(Fe_{1-x}TM_x)_2As_2$ are suppressed in a grossly similar manner by $x$ (regardless of the transition metal) the superconducting domes for Co, Ni and Co/Cu substitutions do not fall onto a single curve.

Another way of parameterizing the effects of TM substitution is to make the assumption that the extra *d*-shell electrons brought along by the dopants become part of the Fermi sea and essentially give rise to a rigid band shift of the density of states at the Fermi level. (Another way of thinking of this is to assume that Co, Ni, and Cu enter into the structure with the same "valence" as Fe and that the extra electrons become part of the delocalized, metallic bonding.) Within this simple model each Co brings an extra electron, each Ni brings two extra electrons and each Cu brings three extra electrons. Making this assumption, a $T - e$ phase diagram can be assembled (34) (Figure 9). This representation of the data shows that the structural / magnetic phase transitions do not scale so well with $e$, but the superconducting domes, specifically on the overdoped side collapse onto a single curve. Although the collapse of the superconducting domes (at least on the overdoped sides) is a noteworthy consequence of this assumption, the validity of parameterizing by something as simply as extra *d* electrons will be discussed further below.

### ***3d* and *4d* doping**

To further explore the $T – x$ and $T – e$ phase diagrams *4d*-TM substitutions were studied for TM = Rh and Pd (66). Figure 8 presents the normalized resistivity for each of these series. When compared to the similar Co and Ni dopings it is clear that there are remarkable similarities between isoeletronic dopant pairs. This observation can be made even more clearly when the $T – x$ phase diagrams for Co and Rh (as well as for Ni and Pd) are plotted together (Figure 9). The Co and Rh $T – x$ phase diagrams are virtually identical, as are the Ni and Pd ones. This is a remarkable result and implies that steric considerations are of minor importance (or somehow, conspiratorially, cancel out) because although Co and Rh (or Ni and Pd) are isoelectronic, they will change the lattice parameters of the parent material in different ways.

The effects of *3d*- and *4d*-TM doping on the lattice parameters of $Ba(Fe_{1-x}TM_x)_2As_2$ are shown in Figure 10. For *3d*-substitution, whereas changes in the *a*-lattice parameter, unit cell volume and the *a*/*c* ratio do not scale well with *x* (65), changes in the *c*-lattice parameter scale do well with *x*. In addition, as shown, changes in the *a*/*c* ratio scale very well with *e*. This would imply that, if only *3d*-doping were to be considered, changes in *x* and the *c*-lattice parameter (or *e* and the *a*/*c* ratio) would be experimentally equivalent variables. On the other hand, the *4d*-doping unambiguously removes this equivalence. Rh and Pd change the *c*-lattice and *a/c* ratio in very different manners than Co and Ni do (66). If the *3d*- and *4d*-TM doping data are considered as a whole, *x* and *e* are not equivalent to any simple combination of the lattice parameters *a* and *c*.

The $T – e$ phase diagram is shown in Figure 11 for TM = Co, Ni, Cu, Rh, Pd, and Co/Cu co-doping over the whole temperature range was well as for an expanded view of the superconducting dome. For each of the doping series the structural / magnetic phase transition can be followed as it is suppressed and its extrapolation goes through the local maximum of related superconducting dome (within the uncertainty of the extrapolation). As can be seen, for the over-doped data there is excellent agreement between the various data sets. For the series that have the structural / magnetic phase transitions suppressed more slowly (as a function of *e*) the samples only superconduct for larger *e*, the "overdoped" region occurs only for larger *e*, and $T_c$ rises to a smaller local maximum. Another way of stating this is that on the underdoped side there is a finite spread between the differing curves, with the series that have the upper structural / magnetic phase transitions being suppressed more rapidly becoming superconducting sooner. When $T_c$ is plotted as a function of $T_s$ and $T_N$ for the underdoped samples there is a clear, essentially linear, correlation (66).

These data suggest that superconductivity can occur in a prescribed region of *e* with $T_c$ depending upon how far $T_s$ or $T_N$ is suppressed on the underdoped side and $T_c$ falling on a universal $T_c(e)$ curve on the overdoped side. This empirical description of the superconducting dome raises several key questions. What defines the lower-*e* limit of the superconducting dome? What it the mechanism / reason for the correlation between suppressing $T_s$ / $T_N$ and increasing $T_c$ on the underdoped side of the dome? And, what is defining the apparently "universal" $T_c(e)$ behavior on the overdoped side of the dome?

It is noteworthy that grossly similar evolution of $T_s$ / $T_N$ and appearance of superconducting dome with TM doping was reported in $Sr(Fe_{1-x}TM_x)_2As_2$ family as well (67-69) including striking similarities between effects of *3d*- and *4d*-TM doping (Co/Rh and Ni/Pd) on $T_c$ (68).

## Electronic changes brought on by TM doping

In order to better characterize the effects of TM substitution on $BaFe_2As_2$, the temperature dependent thermoelectric power (TEP) and Hall resistivity were measured on the $Ba(Fe_{1-x}Co_x)_2As_2$ and $Ba(Fe_{1-x}Cu_x)_2As_2$ series (59). Co- and Cu-doping were chosen because they have the greatest differences in how they affect $T_s$ / $T_N$ and $T_c$: although both Co and Cu suppress (and split) the upper structural and magnetic phase transitions, Co-doping stabilizes superconductivity over a wide range of $e$ – values and Cu-doping does not.

Figure 12 shows TEP and Hall data taken on representative $Ba(Fe_{1-x}Co_x)_2As_2$ and $Ba(Fe_{1-x}Cu_x)_2As_2$ samples. Data from both measurements can be used to further confirm the $T$ - $x$ phase diagrams, with $T_s$, $T_N$ and $T_c$ being readily identifiable (59). The most striking feature, though, is the dramatic change in the TEP between the lower and higher dopings of Co and Cu: the lower doping data falling onto a small-amplitude manifold close to zero and the higher doping data falling onto much larger amplitude manifolds with deep minima at intermediate temperatures. For Co-doping this change takes place for $0.020 < x < 0.024$ and for Cu-doping this change takes place between $0.0077 < x < 0.093$.

The temperature dependent Hall data for the Co- and Cu-doped $BaFe_2As_2$ also show a significant change in behavior between the low and higher doping regimes, with the lower doping samples having low temperature Hall coefficients clustering around – 0.3 nΩ cm/Oe and the higher doping samples having Hall coefficients rapidly rising toward zero. This can be seem most clearly in Figure 19c which presents the $T = 25$ K Hall coefficient data for both doping series plotted as a function of extra electrons, $e$. Both the Co- and the Cu-doped series manifest a sharp increase in the Hall coefficient around $e \sim 0.025$.

For the TEP data, the ratio between the lower and upper limits for the Co- and Cu-doping levels is 2.6, for the Hall coefficient data the two data sets coincide with each other when plotted as a function of $e$ (with Co contributing one extra electron and Cu contributing three extra electrons). Both of these observations are consistent with the hypothesis that Cu adds three times the extra electrons that Co-does when doping $BaFe_2As_2$ at low levels. These observations also indicate that the low side of the superconducting dome in the $T$ – $e$ phase diagrams is apparently delineated by this change in the electronic properties.

Although both the thermoelectric power and Hall coefficient measurements probe convolutions of the Fermi surface/band structural properties and scattering, especially in a multiband intermetallic compound, the dramatic changes seen in the TEP as well as the Hall data are more likely to be associated with changes in the Fermi surface/band structural properties than scattering. This argument is supported by the idea that there may be some form of topological change or a significant distortion in the Fermi surfaces of $Ba(Fe_{1-x}Co_x)_2As_2$ compounds at a given, small change in the *e*-value. In addition, such a sudden change, specifically in the TEP is hard to associate with a change in scattering. At a gross level, drawing on the intuition provided by single-band models, the fact that the change in TEP is so much more dramatic implies that there may be a more dramatic change in the energy derivative of the density of states. Qualitatively similar results on changes in Hall coefficient on Co-doping and on changes in TEP on Ni-doping for a subset of *e* (or *x*) values were reported in references (70, 71) and Ref. (72), respectively.

Angle-resolved photoemission spectroscopy (ARPES) data taken on $Ba(Fe_{1-x}Co_x)_2As_2$ samples (73) support this basic idea. Figure 13a presents Fermi surface mappings of $Ba(Fe_{1-x}Co_x)_2As_2$ for temperatures $T = 13$ K and 150 K. A clear change in the low temperature Fermi surface takes place between $x = 0.024$ and $x = 0.038$. The high temperature Fermi surface changes near $x = 0.024$ as well. This change is quantified further in Figure 13b. These data are consistent with a Lifshitz transition, i.e. with doping, the top of a hole band moves below the Fermi energy. Such a transition could also explain the sudden changes in TEP as well as Hall data.

Figure 13c shows a comparison of the size of the low temperature, Γ and X pockets as inferred from the peak position of the momentum distribution curves for $0.038 \leq x \leq 0.114$. Whereas for underdoped and nearly optimally doped samples the Γ and X pockets are of comparable size and nest fairly well, for $x = 0.114$ the doping has lead to an increasingly poor overlap between these pockets. These data then appear to indicate that the overdoped side of the superconducting dome is governed by relative Fermi surface topology or size, which is controlled by *e*-value in the case of TM doping.

## Jump in specific heat at $T_c$

Due to the complex, and yet well controlled phase diagrams, found for the $Ba(Fe_{1-x}TM_x)_2As_2$ compounds a wide number of studies, detailing the evolution of the superconducting, magnetic and normal state of these materials, have been made. Several properties related to the superconducting state (upper critical field anisotropy (55), penetration depth (74), thermal expansion change at $T_c$ (75)) were reported to show distinct differences for underdoped (orthorhombic, magnetically ordered) and overdoped (tetragonal with no long range magnetic order) sides of the superconducting dome.

On the other hand, the jump in the specific heat at $T_c$ appears not to distinguish between underdoped and overdoped parts of the superconducting dome and, perhaps of more import, shows an unusual dependence on $T_c$ (76) (Figure 14). Not only does this $\Delta C_p \sim T_c^3$ behavior span almost three orders of magnitude between different samples (with $T_c$ values spanning about a decade), but the data for different dopings (both, transition metal on Fe site and K on Ba site) lay on a universal curve with this very simple functional dependence. So far there is no accepted theoretical explanation of this scaling. The possibilities are ranging from the quantum critical normal state nature of the pnictide superconductors (79) to strong “pair-breaking in a broad sense” in superconductors with anisotropic order parameter (79).

## Summary

FeAs based superconductors have been studied extensively over the past two years. During this time of expansive research they have been found to hold great promise, both as potentially technological superconductors, having large and fairly isotropic $H_{c2}$ values even at 20 K, and as new examples of what appears to be novel, meaning not simple electron-phonon-based, superconductivity. Further research (and possibly development) work on these compounds may yield a better understanding of the superconducting state, even in the still mysterious high-$T_c$ copper-oxide materials, and may also produce wires or devices that function at new extremes of temperature and field.

The study of doped $BaFe_2As_2$, in particular $Ba(Fe_{1-x}TM_x)_2As_2$ for TM = Co, Ni, Cu, Rh, Pd, has allowed for a systematic development of clear experimental rules for when and how superconductivity can be induced in this system. Although $BaFe_2As_2$ (as well as other FeAs based superconducting systems) can be tuned into a superconducting state by pressure (hydrostatic and non-hydrostatic) as well as by doping on non-Fe-sites, such as substituting K for Ba, the versatility and variety of TM doping has allowed for extensive thermodynamic, transport, microscopic and spectroscopic measurements. The rules and insight developed from the TM doping studies can be generalized to pressure or isoelectronic substitutions studies by appreciating that band structure can be tuned by steric effects as well as by band filling.

Superconductivity can be induced in the $Ba(Fe_{1-x}TM_x)_2As_2$ materials when (i) the structural / antiferromagnetic phase transitions are suppressed to sufficiently low enough temperatures and (ii) the number of extra electrons added by the TM doping is within a specific window. On the underdoped side of the dome, $T_c$ scales well with the value of the structural and magnetic ordering temperatures ($T_s$ and $T_N$ respectively). On the overdoped side of the dome, $T_c$ scales very well with the number of extra electrons added by the TM-doping.

The lower limit of the superconducting dome is correlated with a dramatic change in the band structure (such as a Lifshitz transition) on the low-*e* side and the offset of the superconducting dome is correlated with an increasingly poor matching of the size of the Γ and X Fermi surface pockets. Once $T_s$ / $T_N$ are suppressed below $T_c$ all of the $T_c - e$ data fall onto a universal curve, implying that the salient physics is controlled by band structure, which for the $Ba(Fe_{1-x}TM_x)_2As_2$ materials is correlated to *e*.

The onset of superconductivity is intimately related to how far $T_s$ / $T_N$ are suppressed. This dramatically illustrated by the correlation between $T_c$ and either of the two upper transition temperatures (66) as well as by the suppressed superconducting dome in $Ba(Fe_{1-x}TM_x)_2As_2$ series that increase $e$ too rapidly compared to how $T_s$ / $T_N$ are suppressed. These correlations imply that there is an antagonistic relation between the superconducting state and the ordered, antiferromagnetic / orthorhombic state. This antagonism is made even clearer by the elastic neutron scattering data shown in Fig. 7 which shows a decrease in the intensity of the magnetic scattering as the sample is cooled into the superconducting state. These observations can be taken as strong evidence of a superconducting pairing mechanism that involves (and requires) magnetic excitations or fluctuations that are present in the tetragonal state, but missing or reduced in the orthorhombic / antiferromagnetic state.

Recently (81) the details of the $T$ - $x$ phase diagram for $Ba(Fe_{1-x}Co_x)_2As_2$ and extensive elastic neutron scattering have been used to demonstrate that these data strongly support the idea that the superconductivity is $s^{\pm}$ in nature. Further insight about the details of superconducting state may be revealed by a theoretical understanding of the $\Delta C_p/T_c$ vs. $T_c$ curve for virtually all superconducting examples of doped $BaFe_2As_2$.

## Final thoughts

At the broadest level the FeAs-based superconductors present a variety of opportunities. By themselves they are a new class of superconductors that (i) have promising properties that may even cause them to become technological materials, and (ii) may present an example of a new type of superconductivity. When taken in the context of the high-$T_c$ oxides, they present the hope that perhaps some clue to exotic superconductivity may be more clearly manifest in the FeAs system and the lessons learned may allow for a better understanding of the long standing question of mechanism and theory for the copper oxide systems. When taken in the context of the past several decades of superconductivity research, they seem to further emphasize that superconductivity is a rather ubiquitous, low temperature ground state for metals and even semimetals. In addition the FeAs materials (as $MgB_2$ before them) point toward compounds that are at the edge between banding and bonding as the promising region of the periodic table for searching for, and hopefully finding, further examples of materials with high $T_c$ values.

## ACKNOWLEDGMENTS
Work at the Ames Laboratory was supported by the US Department of Energy - Basic Energy Sciences under Contract No. DE-AC02-07CH11358. We thank N. Ni and A. Kreyssig for their help with the figures. In addition we would like to thank our colleagues at Ames Laboratory and Iowa State University in collaboration with whom the work presented in this review was done: V.P. Antropov, M.T.C. Apoo, E. Colombier, K.W. Dennis, R.M. Fernandes, A.I. Goldman, R.T. Gordon, B.N. Harmon, A. Kaminski, H. Kim, S. Kim, V.G. Kogan, T. Kondo, A. Kracher, A. Kreyssig, Y.B. Lee, H. Li, C. Liu, C. Martin, R.W. McCallum, R.J. McQueeney, E.D. Mun, S. Nandi, N. Ni, A.D. Palczewski, D.K. Pratt, R. Prozorov, G.D. Samolyuk, J. Schmalian, M.A. Tanatar, A.N. Thaler, M.S. Torikachvili, D. Vaknin, J.Q. Yan.

Figures:

Figure 1: (upper panel) The unit cells of LaFeAsO and $BaFe_2As_2$. Note similar square planar Fe layer (shown as orange balls) capped above and below by As layers (shown as purple balls) in both compounds. (lower panel) Single crystal of $BaFe_2As_2$ grown out of Sn flux shown against a mm - grid.

Figure 2: (a) Anisotropic $H_{c2}(T)$ of a $(Ba_{0.55}K_{0.45})Fe_2As_2$ single crystal (grown out of Sn flux) for applied magnetic field parallel to the *c*-axis and perpendicular to the *c*-axis (parallel to the *ab* plane). Note the exceptionally high value of $H_{c2}$ even at 20 K (after Ref. (19)). (b) Anisotropic $H_{c2}(T)$ for $Ba(Fe_{0.926}Co_{0.074})_2As_2$ (after ref. (55)).

Figure 3: Temperature dependence of the electrical resistivity, anisotropic magnetic susceptibility and specific heat of $BaFe_2As_2$ (left panels) and $CaFe_2As_2$ (right panels). For each compound, all three data sets show the signature of a coupled, structural and antiferromagnetic phase transition. Note that the hysteresis between the warming and cooling data (shown for resistivity) for $CaFe_2As_2$ is further evidence of first order nature of the transitions (after Ref. (17)).

Figure 4: (a) The low temperature, orthorhombic and antiferromagnetically ordered state of $CaFe_2As_2$ that is representative of the low temperature state for $BaFe_2As_2$ and $SrFe_2As_2$ as well (b) The coupling of the structural and antiferromagnetic phase transitions into a strongly first order phase transition is clearly observed in the temperature dependence of the intensity of a magnetic peak and the degree of orthorhombicity (splitting of the *a*- and *b*-lattice parameters) (after (53, 54)).

Figure 5: (upper panel) Normalized in-plane resistivity of $Ba(Fe_{1-x}Co_x)_2As_2$ series; inset: enlarged low temperature part. (lower panel) magnetic susceptibility of $Ba(Fe_{1-x}Co_x)_2As_2$ series ($H \parallel ab$, $H = 1$ T); inserts: low field ($H \parallel ab$, $H = 2.5$ mT) /low temperature zero-field-cooled (ZFC) and field-cooled (FC) magnetic susceptibility and picture of representative $Ba(Fe_{1-x}Co_x)_2As_2$ crystal over a mm grid (after Ref. (55)).

Figure 6: Comparison of the temperature dependent thermodynamic, transport, and scattering data from $Ba(Fe_{0.953}Co_{0.047})_2As_2$: (a) $M(T)/H$ and $dM(T)/HdT$, (b) $R(T)/R_{300K}$ and $dR(T)/R_{300K}dT$, (c) the integrated intensity of the (200) nuclear reflection (circles) and the (½, ½, 1) magnetic reflection (squares). The solid green line shows the power law fit to the higher temperature, magnetic order parameter. The vertical lines through all three panels indicate the structural ($T_s$), magnetic ($T_N$) and superconducting ($T_c$) transition temperatures (after (27)).

Figure 7: $T$ - $x$ phase diagram for $Ba(Fe_{1-x}Co_x)_2As_2$. Solid and open symbols represent orthorhombic and antiferromagnetic transitions respectively. The superconducting transitions are shown by asterisks and half filled squares. As $T_s$ and $T_N$ are suppressed and separate a superconducting dome appears and exists in both the antiferromagnetic / orthorhombic and tetragonal sides of the phase diagram (after (55, 59)).

Figure 8: Temperature dependent, in-plane resistivity (normalized) for Ni-doped, Cu-doped, Cu/Co-doped, Rh-doped and Pd-doped $BaFe_2As_2$ series (after (65, 66)). Insets show low temperature transitions to the superconducting state.

Figure 9: (a) $T$ - $x$ phase diagrams for Co-, Ni-, Cu-, and Cu/Co-dopings of $BaFe_2As_2$. (b) $T$ - $e$ phase diagram for Co-, Ni-, Cu-, and Cu/Co-dopings of $BaFe_2As_2$. The number of extra electrons, $e$, calculated as described in text (after (65)). (c) $T$ - $x$ phase diagrams for Co/Rh doping. (d) $T$ - $x$ diagrams for Ni/Pd doping of $BaFe_2As_2$. Note that isoelectronic doping yields very similar $T$ - $x$ diagrams (after (66)).

Figure 10: (a) Effect of Co-, Ni-, Cu-, and Cu/Co-dopings in $BaFe_2As_2$ on the room temperature c-lattice parameter as a function of x and (b) on the $a/c$ ratio as a function $e$ (after (65)). (c) Effect of Co-, Ni-, Rh-, Pd-dopings in $BaFe_2As_2$ on the $c$-lattice parameter as a function of $x$ and (d) on the $a/c$ ratio as a function of $e$ (after (66)).

Figure 11: (a) $T – e$ phase diagram for Co-, Rh-, Ni-, Pd-, Cu-, and Cu/Co-dopings of $BaFe_2As_2$ (b) enlargement of (a) showing just the superconducting dome (after (65, 66)).

Figure 12: Temperature dependent thermoelectric power for (a) Co-doped and (b) Cu-doped $BaFe_2As_2$. Temperature dependent Hall coefficient for (c) Co-doped and (d) Cu-doped $BaFe_2As_2$. Inset to (d) is a plot of the Hall coefficient at 25 K as a function of extra electron count, $e$, for both Co-doped and Cu-doped $BaFe_2As_2$ (after (59)).

Figure 13: (a) Fermi surface mappings of Co-doped $BaFe_2As_2$ for $T = 13$ K and $T = 150$ K showing the Γ (left) and X (right) pockets. (b) maximum intensity of electrons around the X-pocket for the low temperature data, shown as a function of an alpha angle defined as the angular deviation with respect to the k(110) direction (inset). The intensities are normalized at $\alpha = 90^o$ and symmetrized with respect to $\alpha = 0^o$. (c) Γ- and X-pocket location of the low temperature data extracted via the peak position of the momentum distribution curves for $0.038 \leq x \leq 0.114$. The X-pocket is shifted to the Γ-pocket for easier comparison of their areas (after (73)).

Figure 14: The jump in specific heat associated with the superconducting phase transition plotted as a function of $T_c$ on a *log-log* plot. The dotted line represents a slope of 3, i.e. represents a cubic functional dependence on $T_c$ (after (76) and references therein with additional data from (77, 78)).

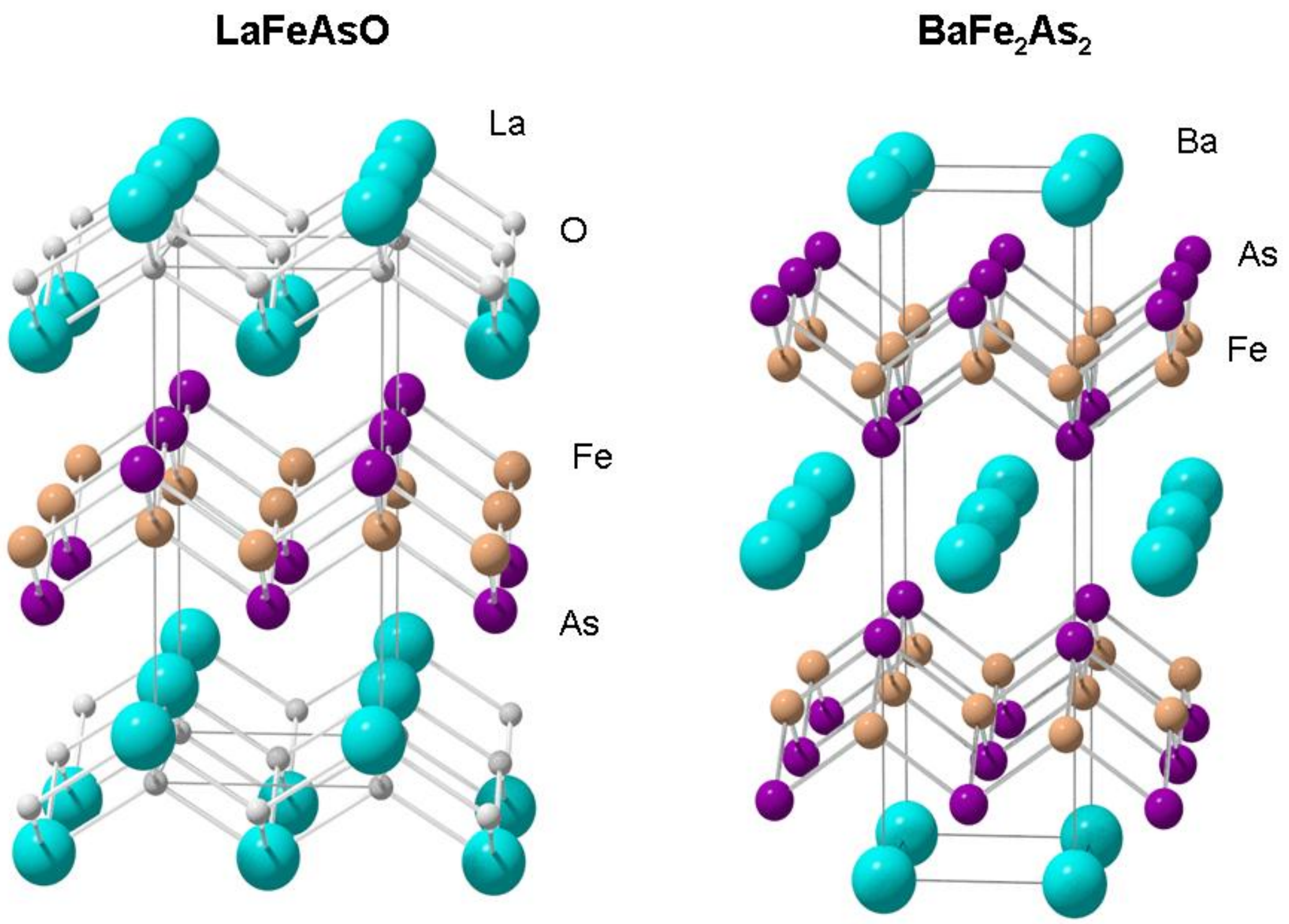

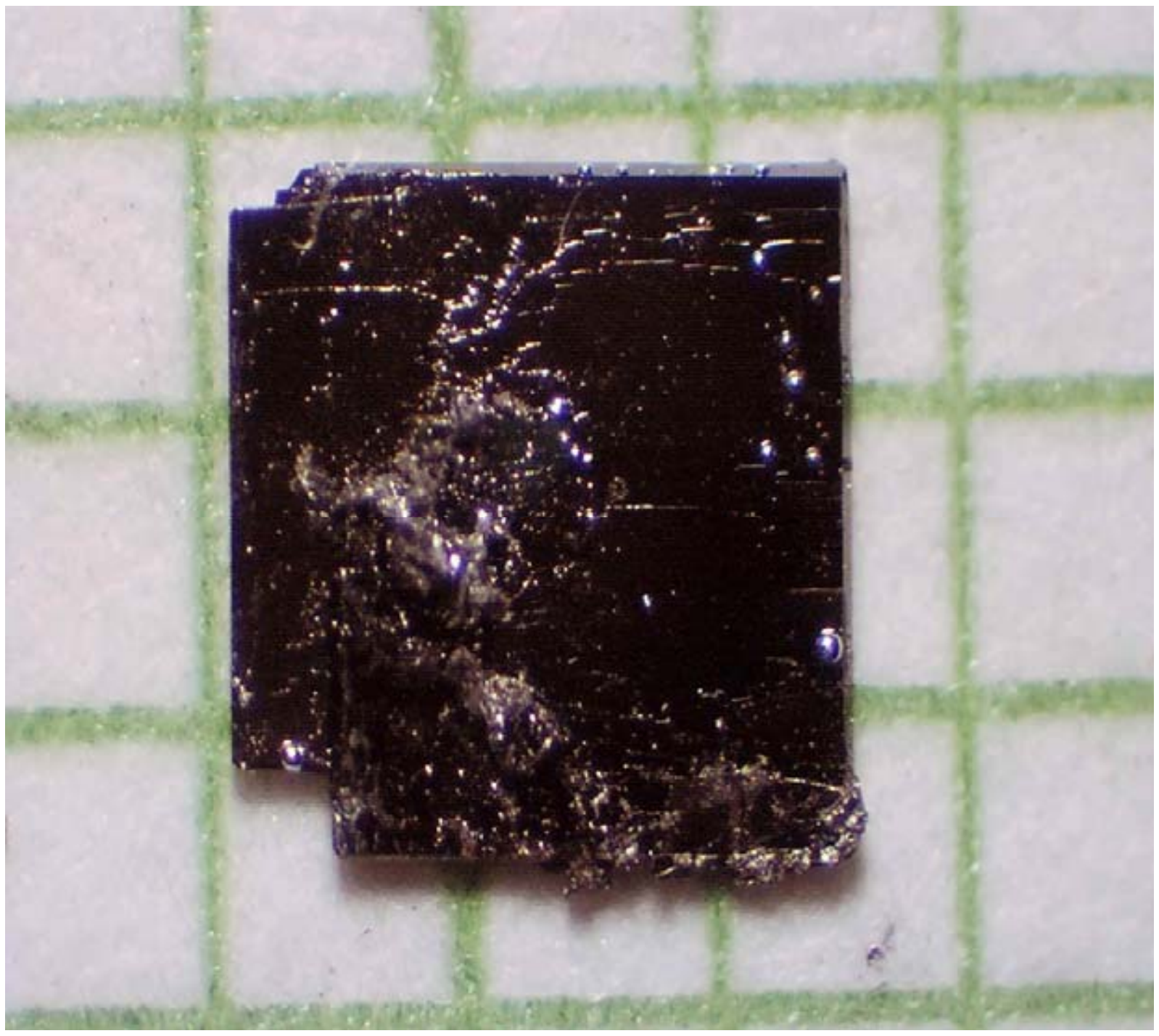

Fig. 1

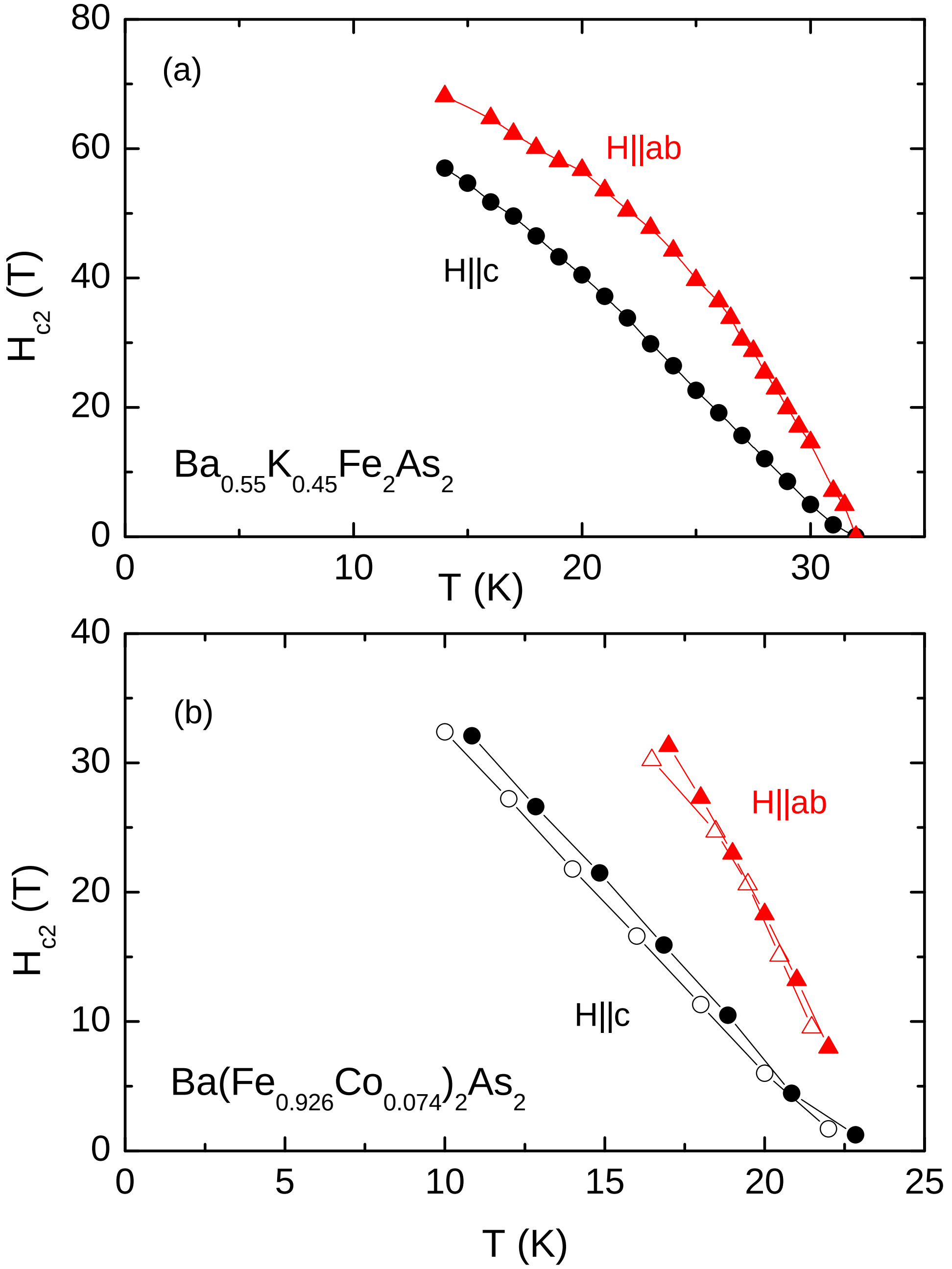

Fig. 2

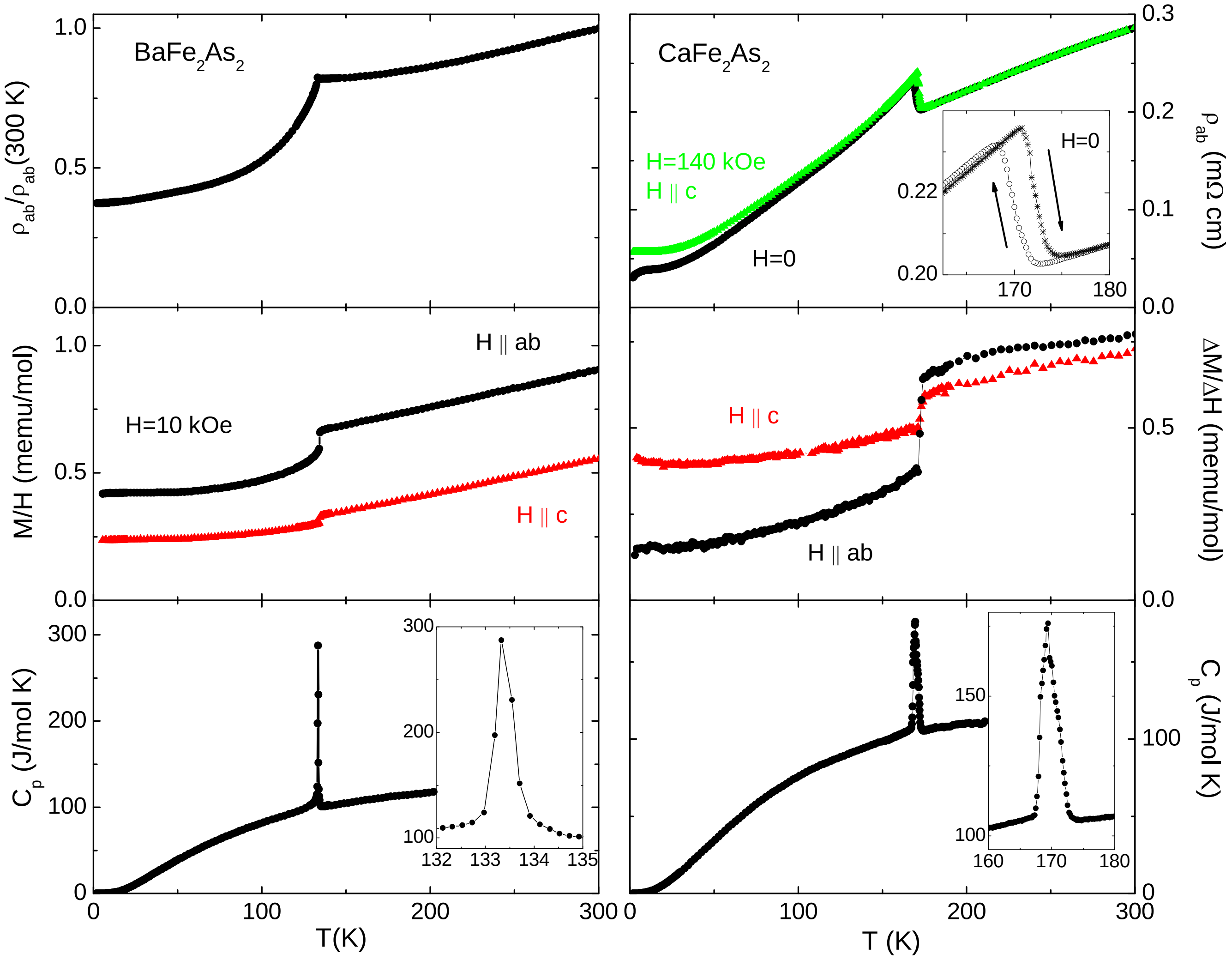

Fig. 3

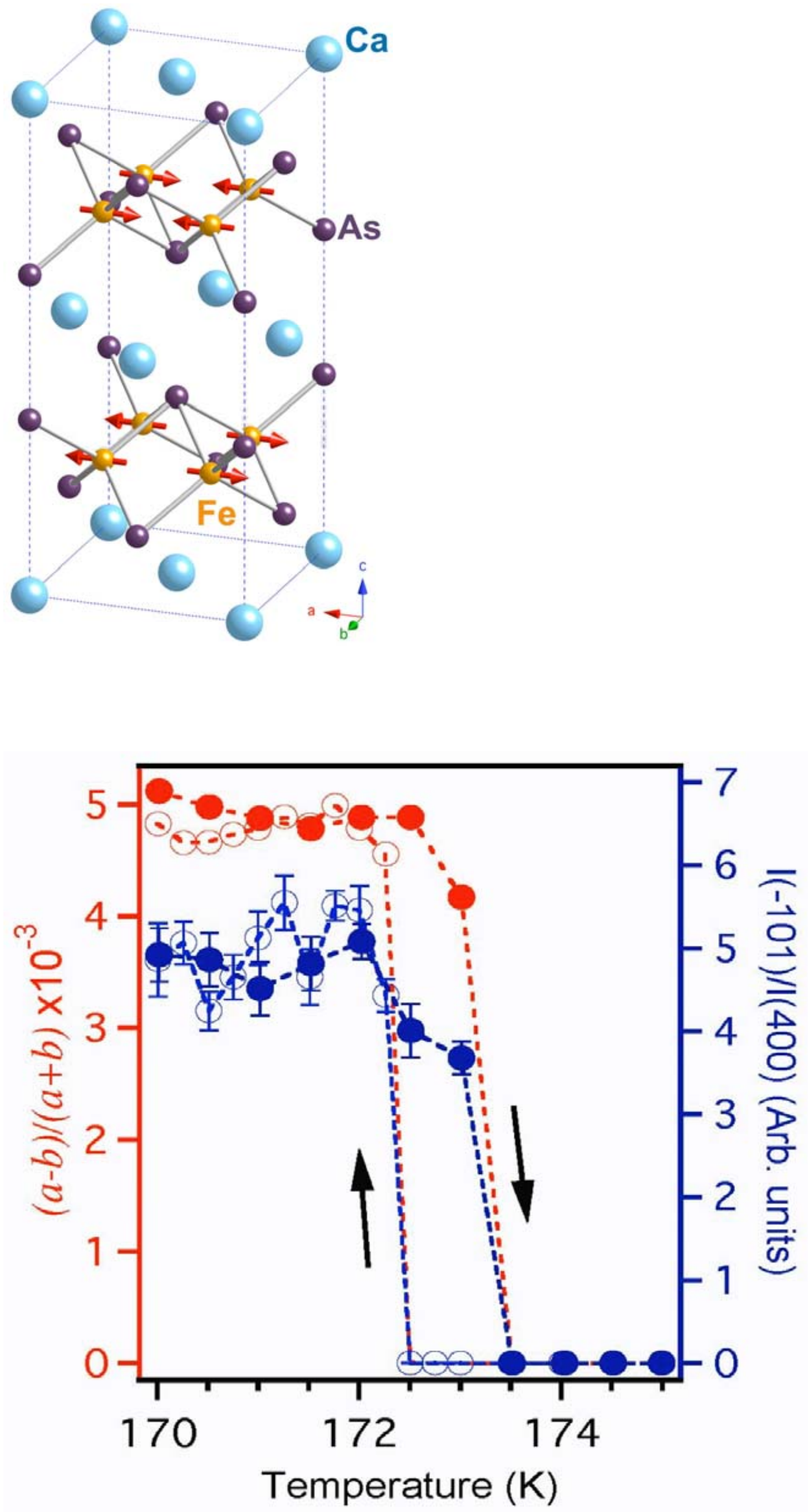

Fig. 4

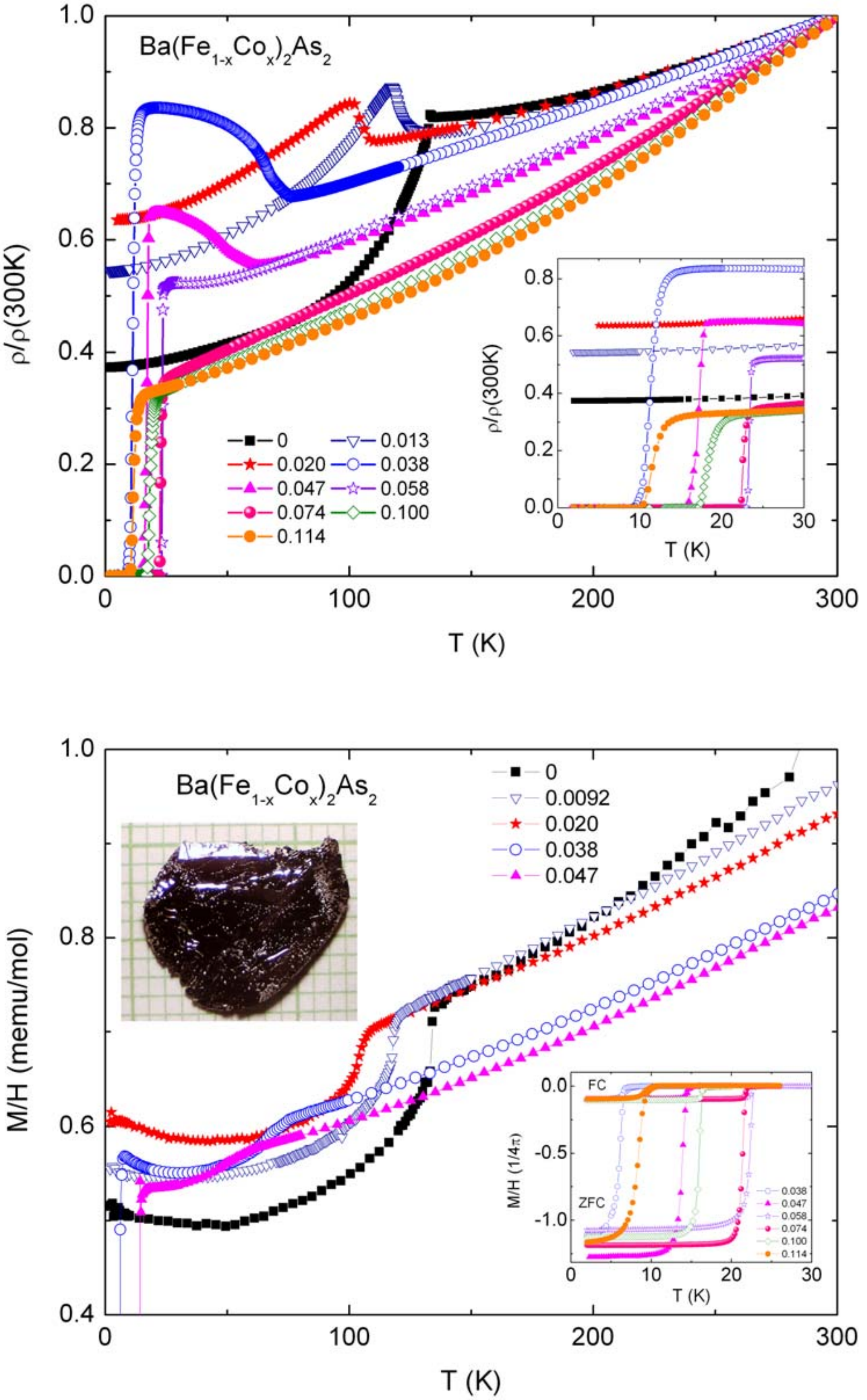

Fig. 5

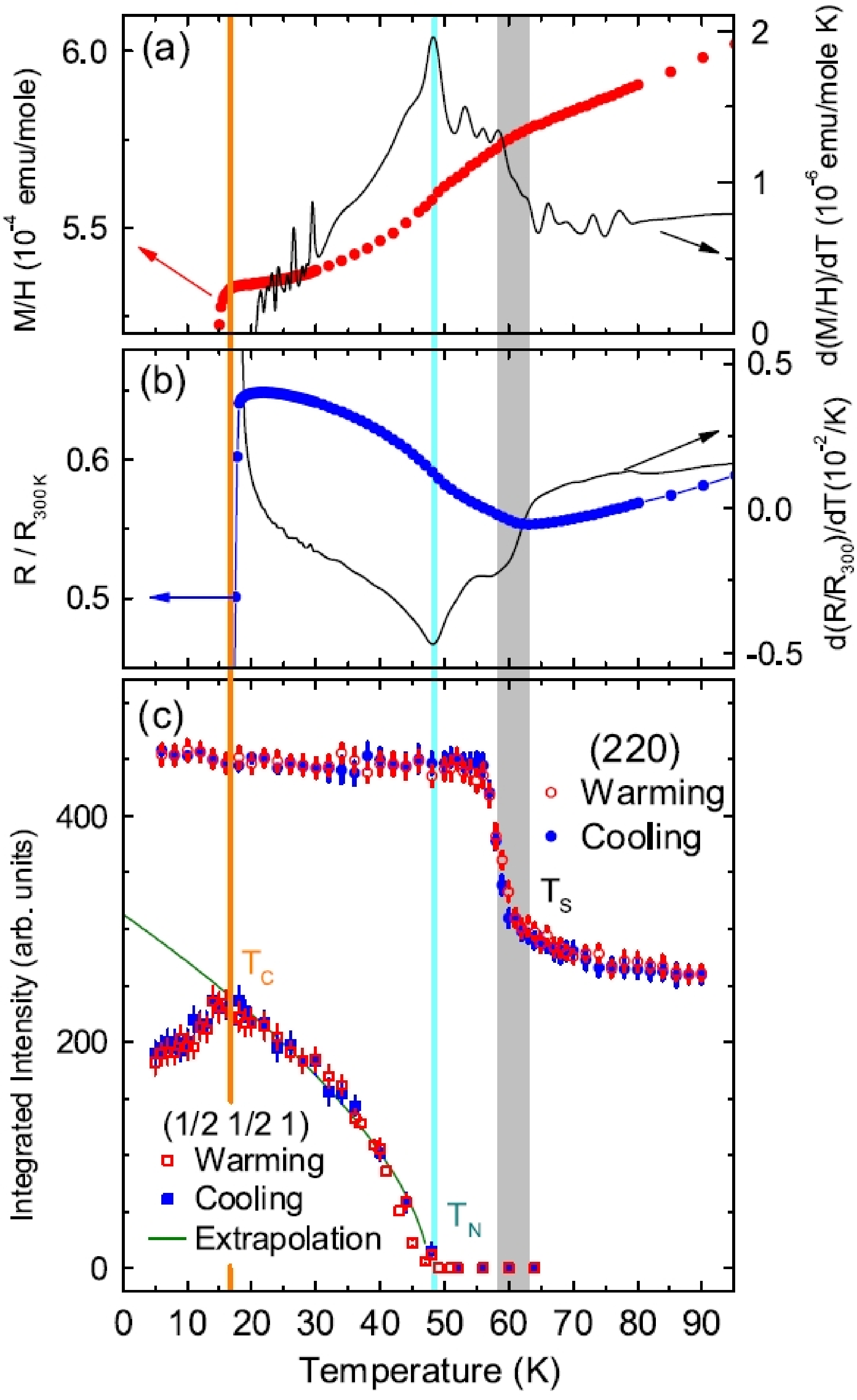

Fig. 6

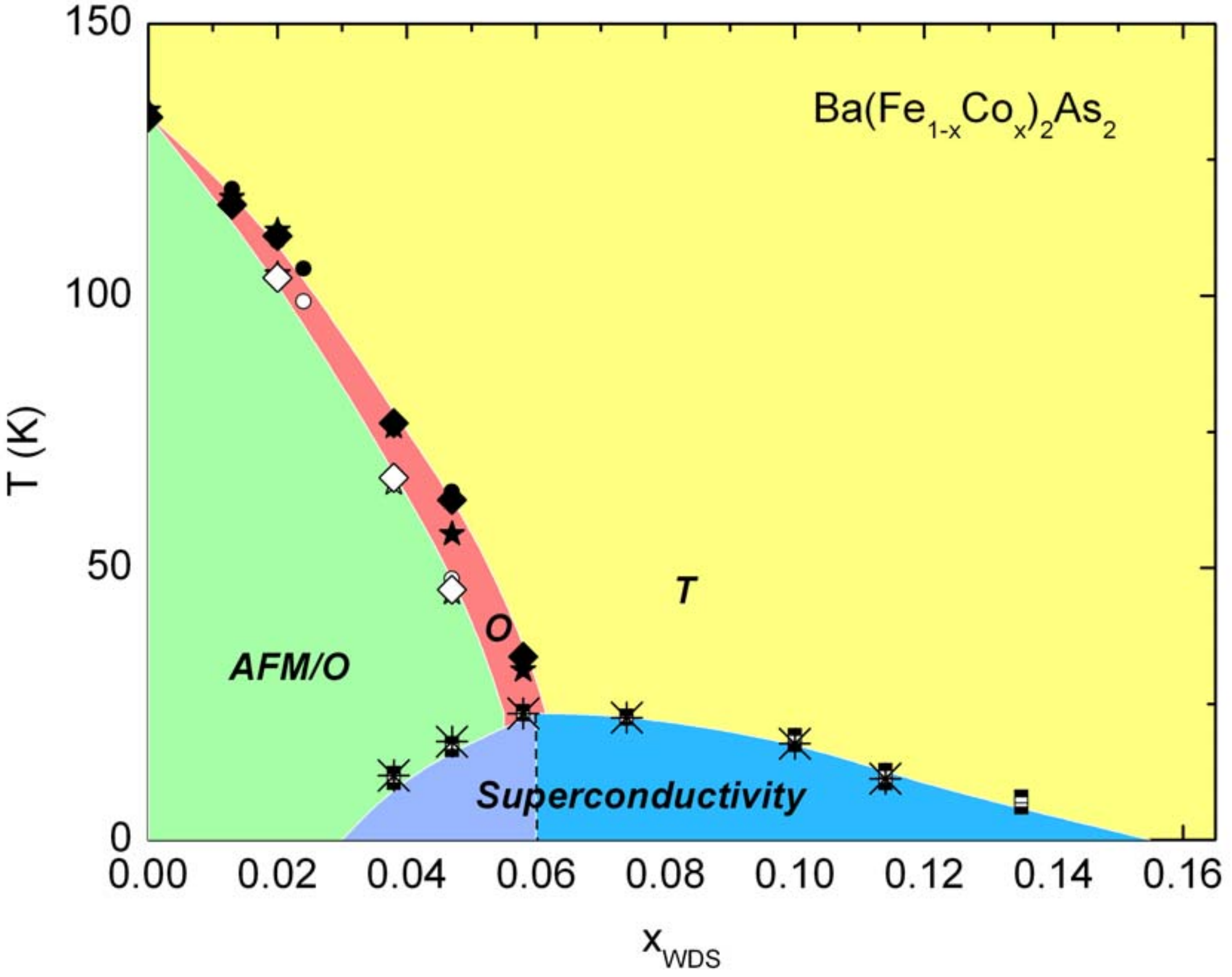

Fig. 7

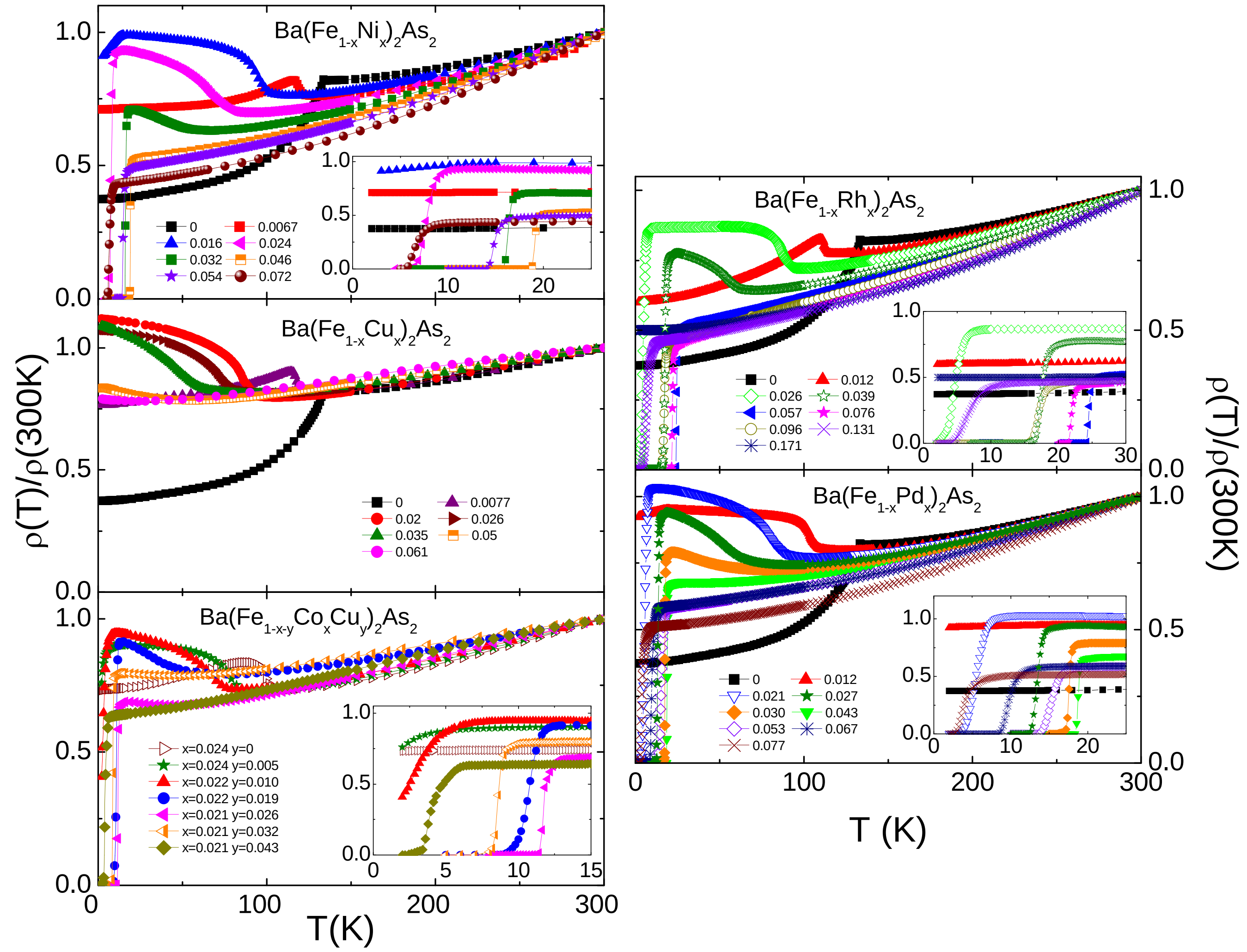

Fig. 8

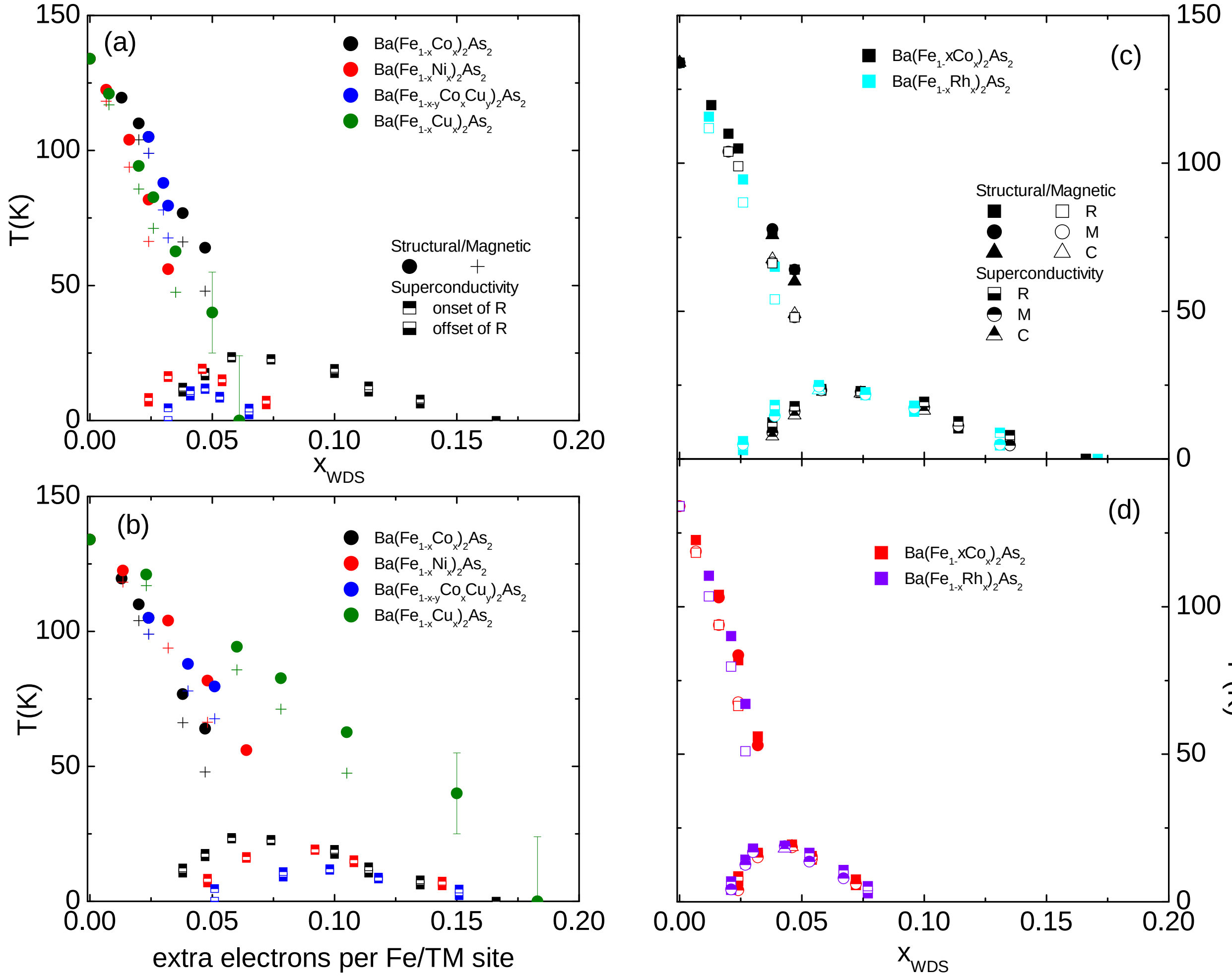

Fig. 9

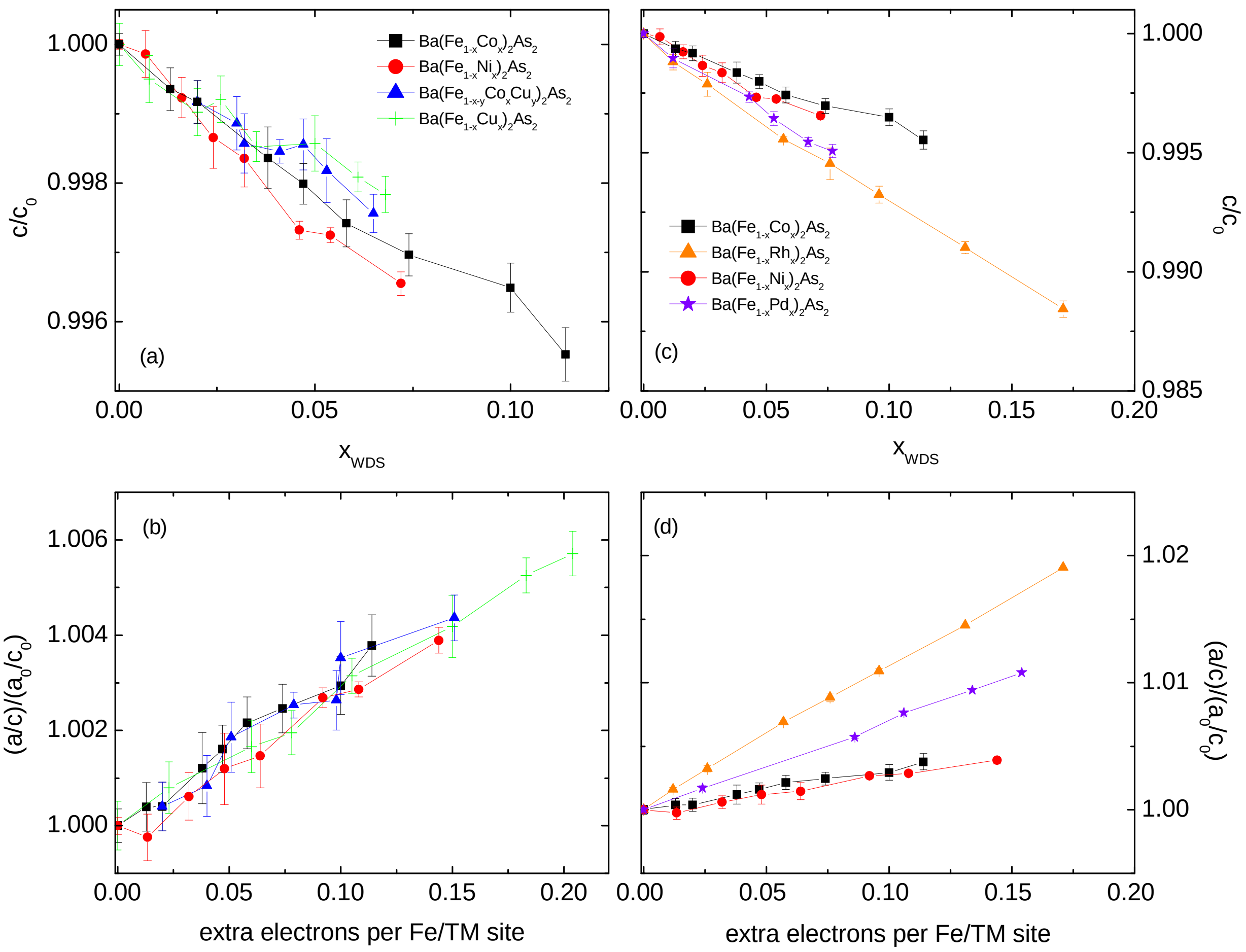

Fig. 10

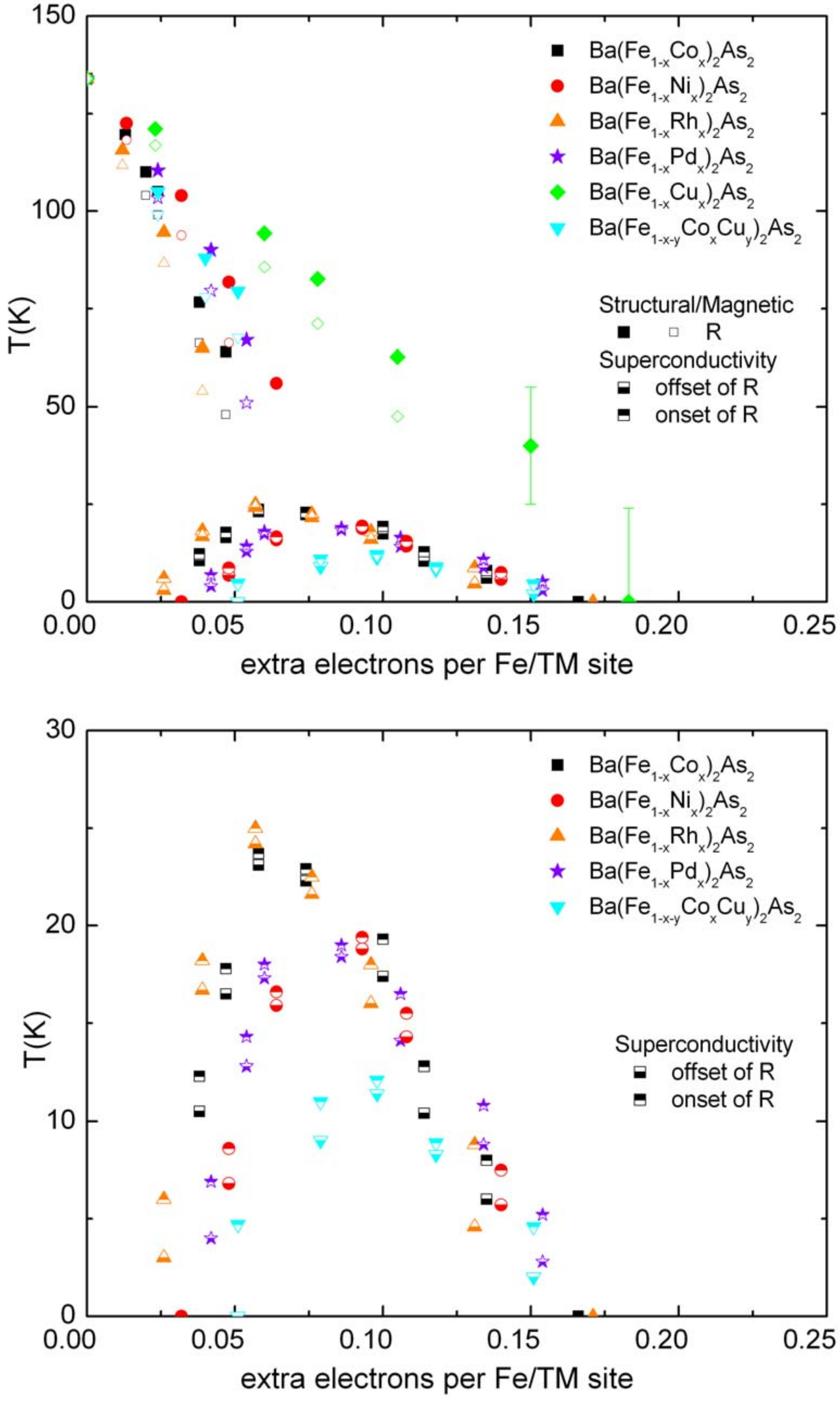

Fig. 11

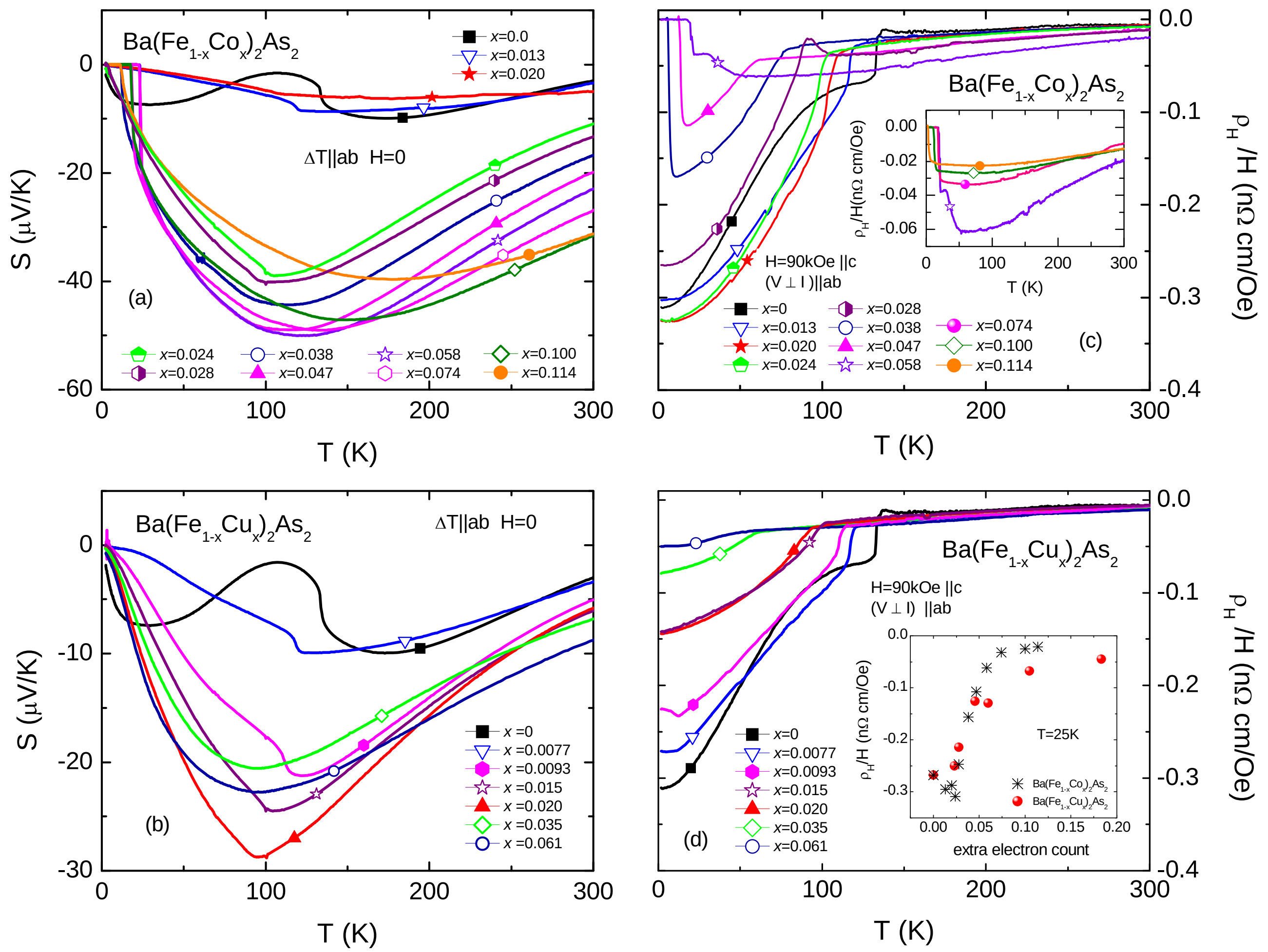

Fig. 12

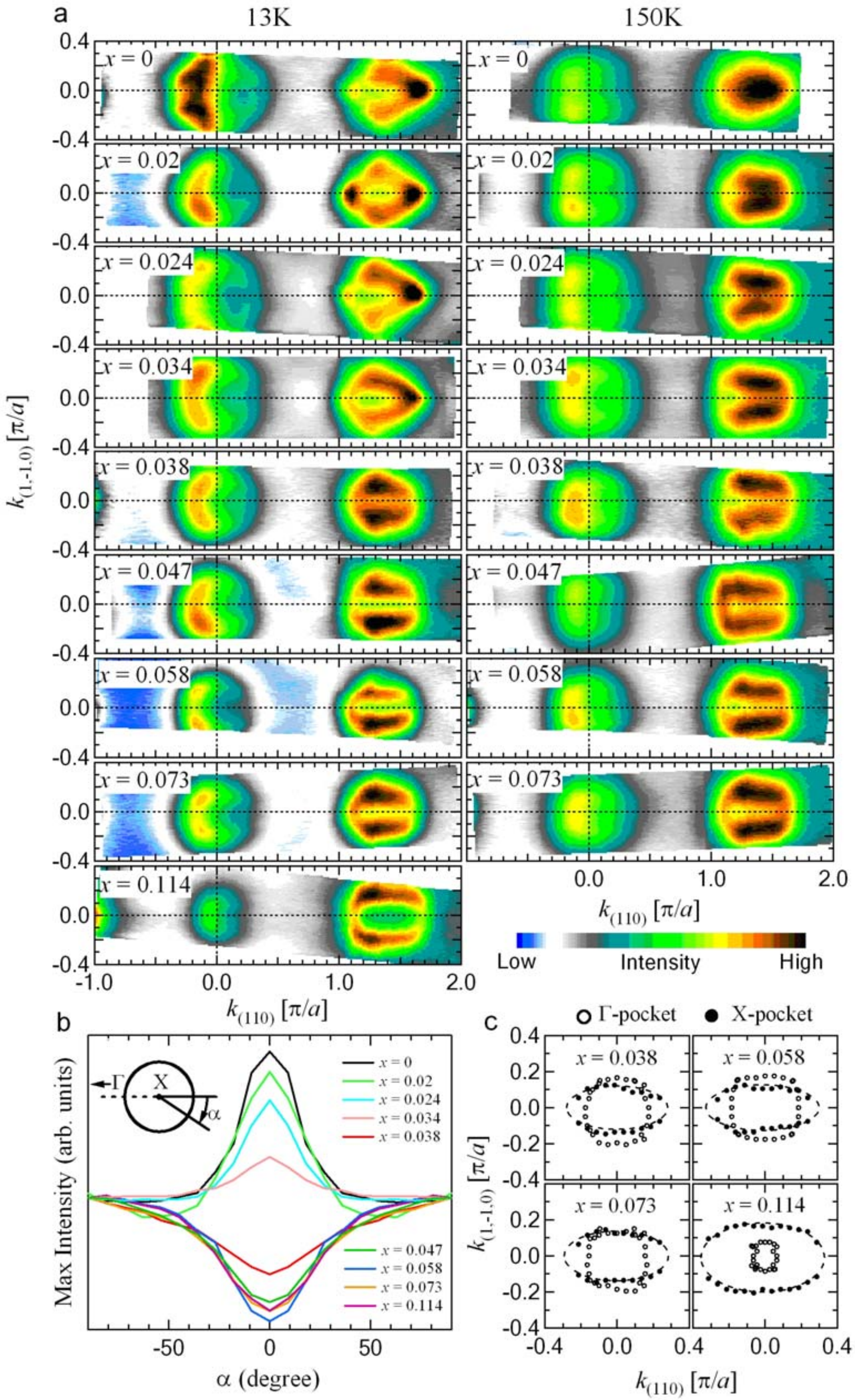

Fig. 13

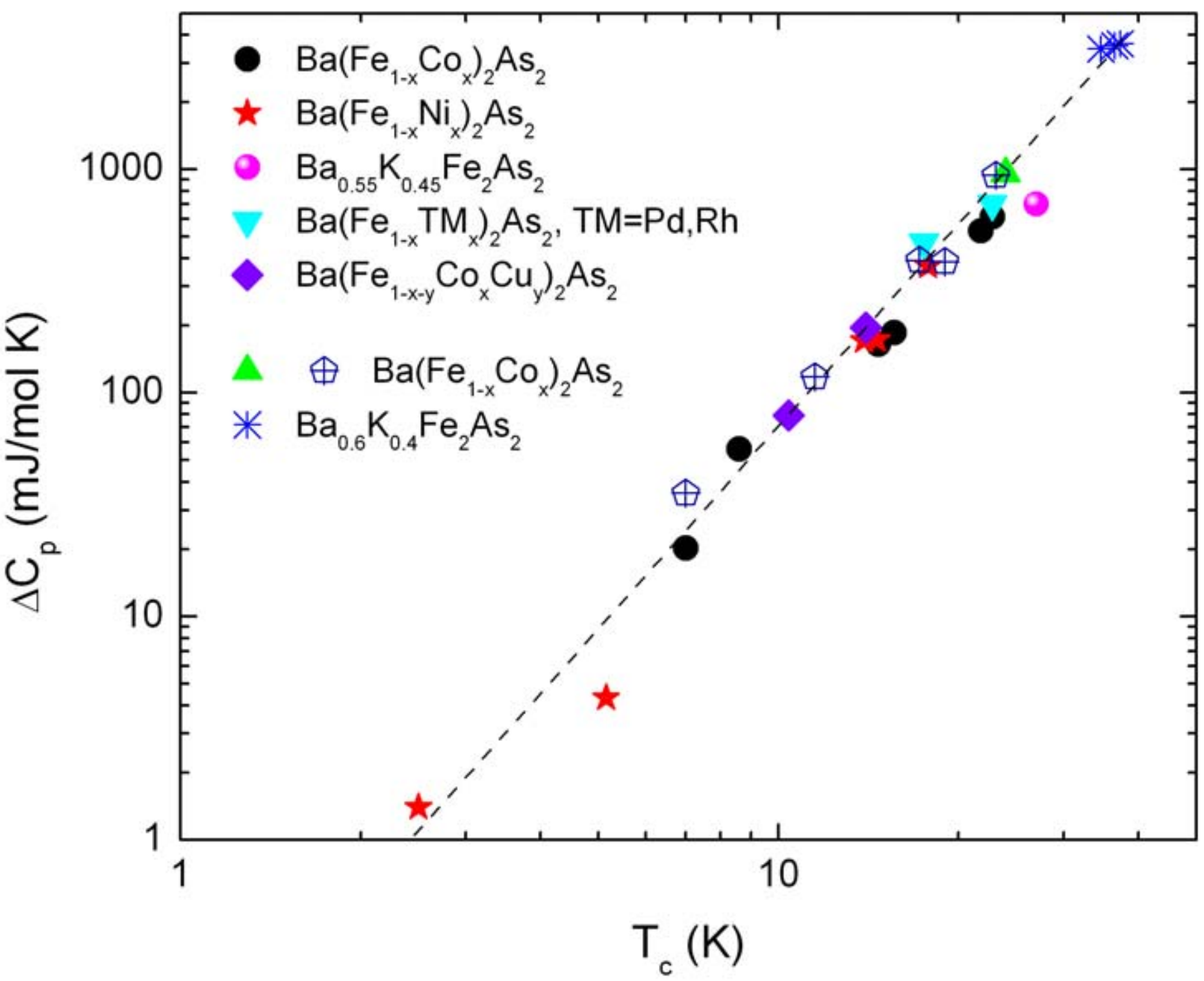

Fig. 14